\documentclass[
12pt,
letterpaper,
reprint,
prl,
aps,
floatfix,
showkeys,
superscriptaddress,
longbibliography,
colorlinks
]{revtex4-2}

\usepackage[utf8]{inputenc}
\usepackage{amsmath,amsthm,amssymb,amsfonts}
\theoremstyle{remark}
\newtheorem{proposition}{\textbf{Proposition}} 
\usepackage{mathtools}
\usepackage{graphicx}
\graphicspath{{./figures/}}
\usepackage{booktabs}
\usepackage{xcolor}
\usepackage{rotating}
\usepackage{orcidlink}
\usepackage{adjustbox}
\usepackage{array}
\usepackage[mathlines]{lineno}
\usepackage{appendix}
\usepackage{aligned-overset}
\usepackage{algorithm}
\usepackage{algpseudocode}
\usepackage{bbm}
\usepackage[normalem]{ulem}
\usepackage{cancel}

\AtBeginDocument{%
  \heavyrulewidth=.08em
  \lightrulewidth=.05em
  \cmidrulewidth=.03em
  \belowrulesep=.65ex
  \belowbottomsep=0pt
  \aboverulesep=.4ex
  \abovetopsep=0pt
  \cmidrulesep=\doublerulesep
  \cmidrulekern=.5em
  \defaultaddspace=.5em
}
\usepackage{hyperref}

\usepackage[framemethod=tikz]{mdframed}
\newmdenv[
  leftmargin=10pt,
  rightmargin=0pt,
  innerleftmargin=10pt,
  innertopmargin=0pt,
  innerbottommargin=0pt,
  skipabove=\topsep,
  skipbelow=\topsep,
  linecolor=gray,
  linewidth=0.5pt,
  frametitleaboveskip=0pt,
  frametitlebelowskip=0pt,
  rightline=false,
  topline=false,
  bottomline=false
]{quotebar}

\newcommand{\abs}[1]{\left\lvert#1\right\rvert} 
\newcommand{\absSmall}[1]{\lvert#1\rvert} 
\newcommand{\floor}[1]{\lfloor#1\rfloor} 
 
\newcommand{\normSmall}[1]{\|#1\|} 
 
\newcommand{\bra}[1]{\langle#1\rvert} 
\newcommand{\ket}[1]{\lvert#1\rangle} 
\newcommand{\proj}[1]{\ket{#1}\bra{#1}} 
 
\newcommand{\braopket}[3]{\langle #1 | #2 | #3\rangle}

\newcommand{\tr}[1]{\text{Tr}\left[#1\right]}
\newcommand{\trSmall}[1]{\text{Tr}[#1]}

\newcommand{\bmatrixByJames}[1]{\left[\;\begin{matrix}#1\end{matrix}\;\right]}

\newcommand{\inth}[1]{\int \text{d}#1\;}

\newcommand{\h}[1]{\hat{#1}}

\newcommand{\C}{\mathbb{C}}

\newcommand{\order}[1]{\mathcal{O}(#1)}

\usepackage{outlines}

\begin{document}
\title{Quantum superresolution and noise spectroscopy with quantum computing}

\author{James W. Gardner\,\orcidlink{0000-0002-8592-1452}}
\email{Contact author: jamesgardner@uchicago.edu}
\affiliation{Pritzker School of Molecular Engineering, University of Chicago, Illinois 60637, USA}
\author{Federico Belliardo\,\orcidlink{0000-0002-1466-396X}}
\affiliation{Pritzker School of Molecular Engineering, University of Chicago, Illinois 60637, USA}
\author{Gideon Lee\,\orcidlink{0000-0003-3306-9189}}
\affiliation{Pritzker School of Molecular Engineering, University of Chicago, Illinois 60637, USA}
\author{Tuvia Gefen\,\orcidlink{0000-0002-3235-4917}}
\affiliation{Racah Institute of Physics, The Hebrew University of Jerusalem, Jerusalem 91904, Givat Ram, Israel}
\author{Liang Jiang\,\orcidlink{0000-0002-0000-9342}\,}
\affiliation{Pritzker School of Molecular Engineering, University of Chicago, Illinois 60637, USA}
\date{\today}

\begin{abstract}
    Quantum metrology of an incoherent signal is a canonical sensing problem related to superresolution and noise spectroscopy. We show that quantum computing can accelerate searches for a weak incoherent signal when the signal and noise are not precisely known. In particular, we consider weak Schur sampling, density matrix exponentiation, and quantum signal processing for testing the rank, purity, and spectral gap of the unknown quantum state to detect the incoherent signal. We show that these algorithms are faster than full-state tomography, which scales with the dimension of the Hilbert space. We apply our results to detecting exoplanets, stochastic gravitational waves, ultralight dark matter, geontropic quantum gravity, and Pauli noise.
\end{abstract}
\maketitle
\allowdisplaybreaks

\begin{figure}
    \centering
    \includegraphics[width=\columnwidth]{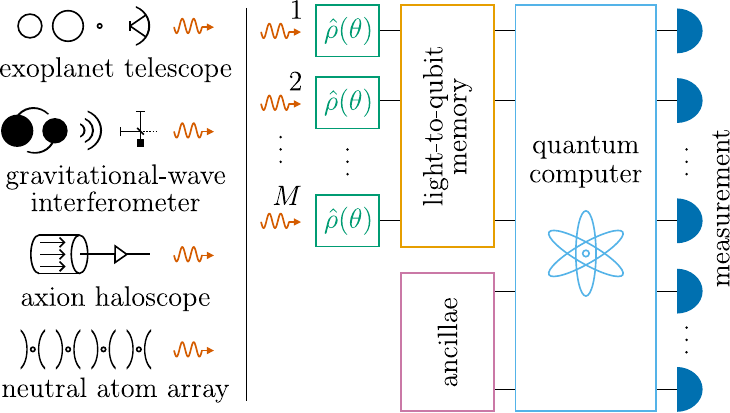}
    \caption{(Left) Sources of weak incoherent signals encoded in bosonic modes. We study unitarily invariant properties of these signals to be robust to nuisance processes such as spatial misalignment. (Right) Circuit of a quantum computer to sense such signals. Many copies of the unknown quantum state are stored in a qubit-based memory, then combined coherently following some algorithm, and finally measured.}
    \label{fig:diagram}
\end{figure}

\textit{Introduction.}---Sensing a weak incoherent signal is a common task at the frontier of physics. Quantum superresolution in optical imaging~\cite{tsang2016quantum,nair2016far,oh2021quantum,PhysRevLett.127.130502,PhysRevA.107.022409} for telescopes~\cite{jha2025multi,kim2025sky} and microscopes~\cite{tenne2019super} enhances the resolving power using spatial-mode demultiplexing (SPADE) beyond that achievable with direct imaging limited by the Rayleigh curse~\cite{rayleigh1880xvii}. Noise spectroscopy~\cite{tsang2012fundamental,ng2016spectrum} for detecting stochastic gravitational waves~\cite{gardner2024stochasticwaveformestimationfundamental,renzini2022stochastic}, ultralight dark matter~\cite{lamoreaux2013analysis,dixit2021searching,shi2023ultimate,shi2025quantum}, and geontropic quantum gravity~\cite{mcculler2022single,vermeulen2025photon} uses number-resolving measurements and other non-Gaussian protocols to accelerate searches for fundamental physics beyond quadrature measurement. Superresolution and noise spectroscopy share the same underlying mathematical structure of overcoming a generalised Rayleigh curse~\cite{tsang2023quantum}. Lindblad detection~\cite{gardner2025lindbladestimationfastprecise} for emerging quantum technologies such as nanoscale nuclear magnetic resonance~\cite{gefen2019overcoming,cohen2020achieving} and qubit spectroscopy~\cite{mouradian2021quantum} has an analogous Rayleigh curse. Improving sensing of incoherent signals therefore has many applications (see Fig.~\ref{fig:diagram}). 

We consider the following canonical incoherent sensing problem. We are given $M$ independent and identically distributed copies of the unknown quantum state:
\begin{align}\label{eq:state}
    \h\rho(\theta)=(1-\theta)\h\rho_n+\theta\h\rho_s
    =\h\rho_n+\theta\h\Delta,\quad\h\Delta=\h\rho_s-\h\rho_n,
\end{align}
where $\h\rho_n$ ($\h\rho_s$) is the noise (signal) state of rank $r_n$ ($r_s$) and $\theta\h \Delta$ is the perturbation. We make no assumptions yet about the structure of $\h\rho_n$ and $\h\rho_s$. We want to decide between the null hypothesis $H_0:\theta=0$ that the classical mixing parameter $\theta$ is zero or the alternative hypothesis $H_1:\theta=\theta_0$ that it equals an \textit{a priori} known, weak value $0<\theta_0\ll1$. (This is a simplification of the more realistic problem where $H_1:\theta\geq\theta_0$.) In the asymptotic limit of many copies $M\gg1$, we want to minimise the worst-case type 2 error $\beta$ provided that the worst-case type 1 error $\alpha\leq\alpha_0$ is held below some fixed value $\alpha_0$. Here, the worst-case errors mean that we maximise them over the respective set of states associated with each hypothesis. 

This problem is well-studied in the case that the noise is known \textit{a priori}. To estimate $\theta$, it is then optimal to project onto the support of $\h\rho_n$, if we know what that support is~\cite{gefen2019overcoming,gorecki2022quantum}. For example, SPADE in optical imaging requires knowing the point spread function of the background. In practice, however, the noise in an experiment is often not exactly known, and thus these existing techniques are not always applicable~\cite{de2025superresolving}. Meanwhile, the sample complexity of full-state tomography scales with the dimension of the Hilbert space~\cite{o2016efficient,haah2016sample}. When features of the noise are unknown, we seek methods with better scaling in the dimension and signal size.

We consider the regime where almost nothing is known about the noise, except for some unitarily invariant property such as its rank, purity, or a spectral gap. These properties only depend on the spectrum of the noise $\h\rho_n$. We assume no \textit{a priori} information about the eigenvectors of the noise $\h\rho_n$, which means that our result is robust to nuisance unitary processes such as spatial misalignment in optical imaging~\cite{de2025superresolving} or desynchronisation (temporal misalignment) in waveform estimation. If the noise was completely unknown, then detecting the signal is not possible, so we aim to study the minimal feasible set of assumptions.

In this work, we solve this incoherent sensing problem using weak Schur sampling, density matrix exponentiation, and quantum signal processing (see Fig.~\ref{fig:diagram}). Using quantum computing algorithms for metrology is an exciting emerging field~\cite{khan2025quantum,allen2025quantum,chin2025quantum,mokeev2025enhancingopticalimagingquantum,de2025superresolving,khan2025quantum1,sarovar2023quantum}. We first review quantum hypothesis testing and these algorithms before discussing our results for rank, purity, and spectral gap testing and their various applications.

\textit{Hypothesis testing.}---We now discuss the quantum limits of hypothesis testing. A loose lower bound on the worst-case type 2 error $\beta$ (and thus on $M$) is given by the quantum simple Chernoff-Stein lemma~\cite{hiai1991proper,ogawa2002strong,brandao2020adversarial}: 
\begin{align}\label{eq:error exponent}
    \beta\geq \exp\left(-M D[\h\rho_n\|\h\rho(\theta_0)]\right)
    ,\quad M \geq \frac{\log(1/\beta)}{D[\h\rho_n\|\h\rho(\theta_0)]},
\end{align}
where the Kullback-Leibler divergence between quantum states is $D(\h\rho\|\h\sigma)=\trSmall{\h\rho(\log(\h\rho)-\log(\h\sigma))}$ and we allow collective measurements on all $M$ copies. Eq.~\ref{eq:error exponent} holds approximately and asymptotically in the following sense of an upper bound on the error exponent:
\begin{align}\label{eq:quantum Chernoff-Stein}
    \lim_{\alpha_0\to 0^+}\liminf_{M\to\infty}\left(\frac{1}{M}\log(1/\beta)\right) \leq D[\h\rho_n\|\h\rho(\theta_0)].
\end{align}
These bounds are loose because they assume that the states are known: they are tight for simple hypotheses but not necessarily for composite hypotheses~\cite{brandao2020adversarial,berta2021composite,lami2025solution,lami2025generalised}.

The Rayleigh curse in optical imaging is that the resolving power of direct imaging vanishes as the separation (signal) goes to zero~\cite{rayleigh1880xvii}. Quantum superresolution methods such as SPADE avoid the Rayleigh curse to achieve better scaling with the separation~\cite{tsang2016quantum,nair2016far,oh2021quantum}. Similarly, we define a generalised Rayleigh curse to be when the error exponent scales quadratically with the signal $\theta_0$ instead of linearly. The following proposition addresses when the Rayleigh curse can be avoided for our sensing problem:
\begin{proposition}\label{prop:kl}
    Let $\h\rho_n$ be a state on a $d$-dimensional Hilbert space $\mathcal{H}=\mathcal{H}_\parallel\oplus\mathcal{H}_\perp$ with $\text{supp}(\h\rho_n)=\mathcal{H}_\parallel$. Consider a support-extending perturbation $\vartheta_0\h\Delta$ with $\text{supp}(\h\rho_n+\vartheta_0\h\Delta)=\mathcal{H}$. We assume that the support-extending component $\h\Delta_\perp:=\h\Pi_\perp\h\Delta\h\Pi_\perp$ is significant such that it controls the perturbation in Frobenius norm, $\normSmall{\h\Delta}=\order{\normSmall{\h\Delta_\perp}}$, where $\h\Pi_\perp$ is the orthogonal projection onto $\mathcal{H}_\perp$. The divergence is then given by:
    \begin{align}\label{eq:Grace+23 Thm.9 modified}
        D(\h\rho_n\|\h\rho_n+\vartheta_0\h\Delta) &= \vartheta_0\,\trSmall{\h\Delta_\perp} + \order{\normSmall{\vartheta_0\h\Delta}^{2}}.
    \end{align}
\end{proposition}
The proof of Proposition~\ref{prop:kl} is given in the Appendix and is similar to the proof of Theorem~9 of Ref.~\cite{grace2022perturbation}. Eq.~\ref{eq:Grace+23 Thm.9 modified} is positive since $\h\Delta_\perp\succ0$ as $\text{supp}(\h\Delta)\supseteq\mathcal{H}_\perp$. 

If the perturbation is instead support-preserving such that $\h\Delta_\perp=0$, then the divergence scales quadratically with $\theta_0$, as we show in the Appendix. The upper bound in Eq.~\ref{eq:quantum Chernoff-Stein} then implies that the Rayleigh curse is unavoidable. A similar result exists for parameter estimation~\cite{gefen2019overcoming,gorecki2022quantum}.

Since we want to avoid the Rayleigh curse, we focus on support-extending perturbations. Proposition~\ref{prop:kl} implies that we may assume that the signal and noise are orthogonal $\trSmall{\h\rho_n\h\rho_s}=0$ without loss of generality. This is because Eq.~\ref{eq:Grace+23 Thm.9 modified} is invariant if we replace $\vartheta_0$ by the reduced signal $\theta_0=\vartheta_0\trSmall{\h\rho_s\h\Pi_\perp}$ and $\h\rho_s$ by the orthogonal state $\h\Pi_\perp\h\rho_s\h\Pi_\perp/\trSmall{\h\rho_s\h\Pi_\perp}$. The state $\h\rho(\theta_0)$ in Eq.~\ref{eq:state} is thus block-diagonal with rank $r=r_n+r_s$, and the noise is not full rank. The divergence in Eq.~\ref{eq:quantum Chernoff-Stein} and sample complexity in Eq.~\ref{eq:error exponent} are then:
\begin{align}\label{eq:quantum Chernoff-Stein, orthogonal}
    D[\h\rho_n\|\h\rho(\theta_0)]=\theta_0+\order{\theta_0^2}, 
    \quad M=\Omega\!\left(\frac{\log(1/\beta)}{\theta_0}\right),
\end{align}
where we used that $\log(1-\theta_0)=-\theta_0+\order{\theta_0^2}$. This lower bound on the sample complexity holds asymptotically in the sense of $\theta_0\ll1$ and the limits in Eq.~\ref{eq:quantum Chernoff-Stein}. The scaling of $M\sim1/\theta_0$ is expected classically for support-extending incoherent signals and should not be confused with the Heisenberg limit which is quantum (see the Appendix).

To compare the performance of different quantum computing algorithms to this lower bound, we use the classical composite Chernoff-Stein lemma~\cite{levitan2002competitive,brandao2020adversarial,lami2025doubly}:
\begin{align}\label{eq:composite Chernoff-Stein} 
    \lim_{\alpha_0\to 0^+}\liminf_{M\to\infty}\left(\frac{1}{M}\log(1/\beta)\right) = \inf_{p\in \mathcal{P},q\in \mathcal{Q}}D(p\|q),
\end{align}
where the Kullback-Leibler divergence between probability distributions of measurement results is $D(p\|q)=\inth{x}p(x)\log[p(x)/q(x)]$. The null hypothesis is $H_0:p\in\mathcal{P}$ and the alternative hypothesis is $H_1:p\in\mathcal{Q}$ given the closed, convex, and disjoint sets $\mathcal{P}$ and $\mathcal{Q}$. The error exponent is the divergence between these sets.

Incoherent sensing problems have largely been studied using the quantum Fisher information (e.g.\ see Refs.~\cite{tsang2016quantum,tsang2012fundamental,tsang2023quantum,gardner2024stochasticwaveformestimationfundamental,gardner2025lindbladestimationfastprecise,de2025superresolving,nair2016far,oh2021quantum,ng2016spectrum,gefen2019overcoming}). The quantum Fisher information, however, is discontinuous if the quantum state is perturbed in trace distance. This makes parameter estimation ill-suited to handling incoherent sensing with quantum computing algorithms, which produce a desired state with high probability up to some small trace distance error. This is why we instead use hypothesis testing.

\textit{Tomography.}---An inefficient way to sense the incoherent signal is full-state tomography which requires~\cite{o2016efficient,haah2016sample}: 
\begin{align}\label{eq:tomography sample complexity}
    M = \mathcal{O}\!\left(\frac{rd\log(1/\beta)}{\theta_0^2}\right).
\end{align}
Tomography gives an estimate of the state which is $\varepsilon$ close in trace distance with probability $1-\beta$ using a collective measurement of the $M$ copies. This implies an estimate of the spectrum which is $\order{\varepsilon}$ close in total variation distance by Weyl's inequality. We can distinguish the signal from zero provided that $\varepsilon\leq\theta_0/8$. Distinguishing the signal from the noise may require additional copies as we will see for spectral gap testing. Full-state tomography is slow since it scales with the dimension $d$, which grows exponentially with system size.

\textit{Weak Schur sampling.}---We now want to find the optimal measurement with sample complexity independent of the dimension. Let $\mathcal{H}=\C^d$ be a $d$-dimensional Hilbert space with $\h\rho^{\otimes M}\in B(\mathcal{H}^{\otimes M})$. The Schur-Weyl duality is that $\mathcal{H}^{\otimes M} \cong \bigoplus_{\vec\lambda} \mathcal{U}_{\vec\lambda,M} \otimes \mathcal{V}_{\vec\lambda,d}$ as representations, where $\vec\lambda$ is a Young diagram (partition of $M$) following the Schur-Weyl distribution, which only depends on $M$ and the spectrum of $\h\rho$~\cite{goodman2000representations,harrow2005applications,christandl2006structure,montanaro2013survey}. $\mathcal{U}_{\vec\lambda,M}$ and $\mathcal{V}_{\vec\lambda,d}$ are the irreducible representations indexed by $\vec\lambda$ of the symmetric group $\text{S}(M)$ and general linear group $\text{GL}(d)$, respectively. 
These two groups arise from permuting the $M$ copies of $\h\rho$ or acting with $\h U^{\otimes M}$, respectively. These two actions commute. 
The spectrum of $\h\rho$ can be estimated by weak Schur sampling (WSS), which projects into the Schur basis to learn just the $\vec{\lambda}$ label, without learning any information about the eigenvectors of $\h\rho$~\cite{keyl2001estimating,childs2007weak,o2015quantum}. If the rank is small $r\ll d$, then this is faster than full-state tomography by a factor of $r/d$. In fact, WSS is a subroutine of full-state tomography and involves coarse-grained versions of the measurements in tomography. WSS is nevertheless faster than tomography if $r\ll d$ since the resulting error for the spectrum in total variation distance is smaller than the error for the state in trace distance.  
Implementing WSS practically is discussed in Refs.~\cite{keyl2001estimating,childs2007weak,o2015quantum,de2025superresolving}. WSS may be performed using a logarithmic working memory of $\order{\log(M)/\log(d)}$ qudits without any loss in sample complexity compared to collective measurement on all $M$ copies~\cite{cervero2023weakschursamplinglogarithmic}. 

For incoherent sensing, we have the following result:
\begin{proposition}\label{prop:wss} 
    To distinguish two unitarily invariant, composite hypotheses $H_0:\h\rho\in\mathcal{S}_0$ and $H_1:\h\rho\in\mathcal{S}_1$, it is optimal in a minimax sense to perform WSS, i.e.\ the worst-case type 1 and type 2 errors are minimal. 
\end{proposition}
The proof of Proposition~\ref{prop:wss} is given in the Appendix and is similar to the proof of Lemma~20 of Ref.~\cite{montanaro2013survey}. The key idea is that any optimal measurement may be replaced by its average over permutations and unitaries, which is a linear combination of projections onto the Schur basis~\cite{childs2007weak}. Proposition~\ref{prop:wss} implies that to determine the optimal sample complexity for incoherent sensing with a unitarily invariant prior, we need to calculate the performance of WSS with logarithmic memory. 

Spectrum estimation may also be performed faster than tomography using only local (i.e.\ single-copy) measurements, but the sample complexity then depends on the dimension $d$ (although tomography has even worse scaling in $d$)~\cite{pelecanos2025beating}. This implies that incoherent sensing with local measurements may be performed faster than tomography. This may be useful if the noise is full rank. We focus instead on the cases which avoid the Rayleigh curse and any scaling in $d$ altogether.

\setlength{\tabcolsep}{6pt}
\begin{table*}[ht!]
\begin{tabular}{@{}lllll@{}}
\toprule
Algorithm  & Sample complexity & Equation & Memory & Background \\ \midrule
Lower bound & $\Omega\!\left(\frac{\log(1/\beta)}{\theta_0}\right)$ & Eq.~\ref{eq:quantum Chernoff-Stein, orthogonal} & $M$ & Refs.~\cite{hiai1991proper,ogawa2002strong,brandao2020adversarial,grace2022perturbation} \\
Full-state tomography & $\mathcal{O}\!\left(\frac{rd\log(1/\beta)}{\theta_0^2}\right)$ & Eq.~\ref{eq:tomography sample complexity} & $M$ & Refs.~\cite{o2016efficient,haah2016sample} \\
Rank testing (WSS) & $\Theta\!\left(\frac{r_n^2\log(1/\beta)}{\theta_0}\right)$ & Eq.~\ref{eq:rank testing, sample complexity} & $\mathcal{O}[\log(M)]$ & Ref.~\cite{o2015quantum} \\
Purity testing (WSS) & $\Theta\!\left(\frac{\log(1/\beta)}{\theta_0}\right)$ & Eq.~\ref{eq:rank testing, sample complexity} & $\mathcal{O}[\log(M)]$ & Ref.~\cite{o2015quantum} \\
Spectral gap testing (WSS) & $\mathcal{O}\!\left(\frac{r^2\log(1/\beta)}{\min(\theta_0^2,\Lambda^2)}\right)$  & Eq.~\ref{eq:spectral gap testing, sample complexity} & $\mathcal{O}[\log(M)]$ & Ref.~\cite{o2016efficient} \\
Spectral gap testing (WSS), $\Lambda^2\gg\theta_0$ & $\mathcal{O}\!\left(\frac{r_n^2\log(1/\beta)}{\theta_0}\right)$ & Eq.~\ref{eq:Tuvia's hybrid scheme} & $\mathcal{O}[\log(M)]$ & Refs.~\cite{o2015quantum,o2016efficient} \\
Spectral gap testing (DME-QSP) & $\mathcal{O}\!\left(\frac{\log(1/\delta)^2\log(1/\beta)}{\Lambda^2 \varepsilon \theta_0}\right)$ &  Eq.~\ref{eq:QSP sample complexity} & $2$  & Refs.~\cite{lloyd2014quantum,low2017quantum,martyn2021grand,low2017hamiltonian,low2017optimal,motlagh2024generalized,wang2024resource,mokeev2025enhancingopticalimagingquantum} \\
\bottomrule
\end{tabular}%
\caption{Sample complexity and memory requirements of algorithms for incoherent sensing. Here, $M$ is the number of copies, $\theta_0$ is the signal, $\beta$ is the worst-case type 2 error, $d$ is the dimension, $r$ is the total rank, $r_n$ is the rank of the noise, $\Lambda$ is the spectral gap, $\varepsilon$ is the DME-QSP trace distance error, and $\delta$ is the DME-QSP failure probability.}
\label{tab:sample complexity}
\end{table*}

\textit{Rank and purity testing.}---We now determine the performance of WSS for incoherent sensing given different unitarily invariant priors to see whether WSS avoids the Rayleigh curse, which we have shown to be possible for support-extending perturbations. We start with rank testing, where we assume that we know the rank of the noise $r_n$ in Eq.~\ref{eq:state} and want to test whether the state has higher rank. Theorem 1.11 of Ref.~\cite{o2015quantum} implies that rank testing for incoherent sensing requires:
\begin{align}\label{eq:rank testing, sample complexity}
    M=\Theta\!\left(\frac{r_n^2\log(1/\beta)}{\theta_0}\right),
\end{align}
for type 2 error $\beta$ and zero type 1 error $\alpha=0$ (i.e.\ a one-sided error). Here, the state is either rank $r_n$ or $\theta_0$ far in trace distance from having rank $r_n$. The rank testing algorithm detects the signal if the length of the measured Young diagram is greater than $r_n$. This shows that WSS avoids the Rayleigh curse. The lower bound on $M$ in Eq.~\ref{eq:quantum Chernoff-Stein, orthogonal} is loose because it misses the $r_n^2$ scaling in Eq.~\ref{eq:rank testing, sample complexity}, as we summarise in Table~\ref{tab:sample complexity}. The exception is purity testing, where $r_n=1$ and we test whether the state is pure or mixed. For the qubit model of $r_n=r_s=1$, WSS with logarithmic memory is faster than a swap test, which requires only two-copy memory~\cite{buadescu2019proceedings,fanizza2020beyond,len2022multiparameter,de2025superresolving}. The swap test, in fact, is the two-copy version of WSS.

\textit{Spectral gap testing.}---We now apply WSS to a different unitarily invariant prior. Suppose that we only know that the nonzero eigenvalues of the noise $\h\rho_n$ lie above $(\theta_0+\Lambda)/(1-\theta_0)$ for some spectral gap $\Lambda>0$. The gap ensures that the eigenvalues of $(1-\theta_0)\h\rho_n$ and $\theta_0\h\rho_s$ in Eq.~\ref{eq:state} are separated by $\Lambda$. Although we then know that $r_n\leq \floor{(1-\theta_0)/(\theta_0+\Lambda)}$ by normalisation of $\h\rho_n$, performing rank testing will miss signals of lower rank, so we need a different approach. 

To estimate the spectrum of $\h\rho(\theta)$ to within $\varepsilon$ in total variation distance with probability $1-\delta$ with WSS requires a sample complexity of $M=\order{r^2\log(1/\delta)/\varepsilon^2}$~\cite{o2016efficient}. To distinguish the hypotheses with classical postprocessing, we need to distinguish the signal from the noise and zero such that $\varepsilon=\min(\theta_0,\Lambda)/8$ is sufficient (also for tomography). Here, we assume $r_s=1$ such that $r=r_n+1$ for simplicity. 
The remaining error in the hypothesis test comes from the failure probability $\beta\sim\delta$ such that the sample complexity is:
\begin{align}\label{eq:spectral gap testing, sample complexity}
    M=\mathcal{O}\!\left(\frac{r^2\log(1/\beta)}{\min(\theta_0^2,\Lambda^2)}\right),    
\end{align}
treating the type 1 and type 2 errors symmetrically. This does not attain the lower bound on $M$ in Eq.~\ref{eq:quantum Chernoff-Stein, orthogonal} and is Rayleigh cursed but independent of the dimension. 

If the spectral gap is larger than the signal $\Lambda^2\gg\theta_0$, then the Rayleigh curse can be avoided. We estimate the spectrum with $\varepsilon=\Lambda/8$ to determine the rank of the noise $r_n$, then perform rank testing. The sample complexity is:
\begin{align}\label{eq:Tuvia's hybrid scheme}
    M=\mathcal{O}\!\left(\frac{r^2\log(1/\beta)}{\Lambda^2}\right)+\mathcal{O}\!\left(\frac{r_n^2\log(1/\beta)}{\theta_0}\right),
\end{align}
where the second term dominates if $\Lambda^2\gg\theta_0$. This approximately attains the lower bound in Eq.~\ref{eq:quantum Chernoff-Stein, orthogonal} up to a factor of $r_n^2$ and constants.

\textit{Constant memory.}---The sample complexity above requires logarithmic memory. Constant memory is more practical. We show that density matrix exponentiation (DME)~\cite{lloyd2014quantum} and quantum signal processing (QSP)~\cite{low2017quantum,martyn2021grand,low2017hamiltonian,low2017optimal,motlagh2024generalized,wang2024resource} can perform spectral gap testing with two-copy memory and sample complexity independent of the dimension. Our goal is to conditionally block-diagonalise the state:
\begin{align}\label{eq:desired state}
    \mathcal{E}[\h\rho(\theta)\otimes\proj{0}]=(1-\theta)\h\rho_n\otimes\proj{0}+\theta\h\rho_s\otimes\proj{1},
\end{align}
such that measuring the ancilla in $\ket{0},\ket{1}$ yields a Bernoulli random variable with parameter $\theta$. DME-QSP can approximate the channel above to a trace distance $\varepsilon$ and failure probability $\delta$. We proceed in two stages: DME to construct the controlled unitary $e^{\pm i \h\rho(\theta) x}$ for some small $x$ and QSP where successive uses of DME are interleaved with rotations. DME can be approximated within trace distance $\order{x^2}$ using $\order{1}$ copies in two-copy memory using infinitesimal swap operations~\cite{lloyd2014quantum}. $\mathcal{O}[\log(1/\delta)/(\Lambda x)]$ applications of DME can approximate the Heaviside step function to within $\delta$ uniformly outside the spectral gap~\cite{low2017quantum,martyn2021grand,low2017hamiltonian,low2017optimal,motlagh2024generalized,wang2024resource}. The sample complexity of DME-QSP is therefore $M_\text{qsp}=\mathcal{O}[\log(1/\delta)^2/(\Lambda^2 \varepsilon)]$ with trace distance error $\varepsilon=\mathcal{O}[\log(1/\delta)x/\Lambda]$~\cite{mokeev2025enhancingopticalimagingquantum}. Measuring the ancilla in $\ket{0},\ket{1}$ then yields a Bernoulli random variable with parameter $(1-2\delta)\theta+\delta+\order{\varepsilon}$. The number of Bernoulli measurements is $M_\text{ber}\approx\log(1/\beta)/\theta_0$ such that the total sample complexity is:
\begin{align}\label{eq:QSP sample complexity}
    M_\text{tot}=M_\text{ber}M_\text{qsp}\approx\mathcal{O}\!\left(\frac{\log(1/\delta)^2\log(1/\beta)}{\Lambda^2 \varepsilon \theta_0}\right),
\end{align}
which is independent of the dimension. We prove this in the Appendix. DME-QSP is slower than WSS since $\varepsilon,\delta\ll\theta_0$ and $r_n\leq\floor{(1-\theta_0)/(\theta_0+\Lambda)}$. Tomography of the conditional state after the ancilla measurement could be used to learn the noise.

\textit{Impact of noise.}---We now consider depolarising noise $(1-\gamma)\h\rho(\theta)+\gamma\h I/d$, where $\h I$ is the identity and $\h I/d$ is the maximally mixed state. If the signal dominates the noise $\gamma\ll\theta_0\ll1$, then the divergence $D\approx(1-\gamma)\theta_0+\frac{1}{2}\theta_0^2$ scales linearly with the signal. If instead the noise is dominant $\gamma\gg\theta_0$, then the divergence $D\approx\frac{1}{2}\theta_0^2\mathcal{I}_C$ scales quadratically with the signal, where $\mathcal{I}_C$ is the classical Fisher information at $\theta=0$ which depends on $\gamma$, $d$, and the spectra of $\h\rho_n,\h\rho_s$. The Rayleigh curse is thus unavoidable by Eq.~\ref{eq:quantum Chernoff-Stein}. We now focus on spectral gap testing. The full-rank state means that full-state tomography and spectrum estimation both require~\cite{o2016efficient,haah2016sample}:
\begin{align}
    M=\mathcal{O}\!\left(\frac{d^2\log(1/\beta)}{(1-\gamma)^2\min(\theta_0^2,\Lambda^2)}\right),
\end{align}
where $\varepsilon=(1-\gamma)\min(\theta_0,\Lambda)/8$. In comparison, DME-QSP remains independent of the dimension, although it still exhibits the Rayleigh curse. (WSS is still optimal, but the spectrum estimation bound is loose.) If the noise is subdominant $\gamma\ll\theta_0$, then the DME-QSP error exponent in Eq.~\ref{eq:error exponent after QSP, composite} is perturbed by $\delta\to\delta+\gamma$, dropping factors of $r_n/d\ll1$. If instead the noise is dominant $\gamma\gg\theta_0$, yet there remains a spectral gap, then the error exponent scales as $\theta_0^2$.

\textit{Exoplanet detection.}---We now discuss applications of incoherent sensing, starting with exoplanet detection. WSS for optical imaging was proposed in Ref.~\cite{de2025superresolving} but studied only for the qubit model of $r_n=r_s=1$ corresponding after heralding to pure sources of a single star and planet. We show that WSS is useful more generally: it can be applied to mixed sources, multi-star systems, and given only spectral gap information~\cite{thorsett1993psr,thebault2025complete}. We also use hypothesis testing to show that DME-QSP works with constant memory.

\textit{Lindblad detection.}---We now consider Lindblad jump operator detection. The Lindblad master equation is:
\begin{align}
    \dot{\h\rho} = -i[\h H, \h\rho] + {\textstyle \sum_{j=1}^{N+S}} \kappa_j \mathcal{L}_{\h L_j}(\h \rho),
\end{align}
where $\mathcal{L}_{\h L_j}(\h \rho)=\h L_j\h\rho\h L_j^\dagger-\frac{1}{2}\{\h L_j^\dagger\h L_j,\h\rho\}$. In the short-time limit, the final state is $\h\rho_i+t\dot{\h\rho}_i$. If we want to sense $S$ weak Lindblad operators in the presence of the other $N$, the reduced signal is $\theta_0= \sum_{j=N+1}^{N+S} \kappa_j t\, \trSmall{\h\Pi_\perp\mathcal{L}_{\h L_j}(\h \rho)}$. If $\h\rho_i$ is pure and we know $N$, then we can perform rank testing. This could be applied to detect correlated dephasing or decay in the presence of unknown Pauli noise foregrounds~\cite{shaw2024multi,mok2024universal,ferioli2023non,mouradian2021quantum,gefen2019overcoming,cohen2020achieving}. The generalisation to Kraus operator detection is given in the Appendix.

\textit{Noise spectroscopy.}---We now consider estimating the power spectral density of a classical signal coupled to a linear system. (This noise spectroscopy problem of sensing a stochastic waveform should be distinguished from estimating a deterministic waveform~\cite{Tsang+2011,PhysRevLett.132.130801,ding2026holevo}.) The state is finite-dimensional given a finite time-bandwidth product and energy cutoff. Ref.~\cite{gardner2024stochasticwaveformestimationfundamental} showed that it is optimal to prepare the bosonic mode at a given frequency in a Gottesman-Kitaev-Preskill state $\h\varrho$~\cite{gottesman2000encoding}. Suppose that we have unknown state preparation errors $\h\rho=\Xi(\h\varrho)$ and only know the rank or spectral gap of $\h\rho$, then we can perform WSS or DME-QSP to detect the signal. The reduced signal is $\theta_0=\kappa t\,\trSmall{\h\Pi_\perp\mathcal{L}_{\h x}(\h\rho)}$ for tuned interferometry for detecting stochastic gravitational waves or geontropic quantum gravity and $\theta_0=\kappa t\,\text{Tr}(\h\Pi_\perp[\mathcal{L}_{\h x}(\h\rho)+\mathcal{L}_{\h p}(\h\rho)])$ for ultralight dark matter~\cite{gardner2024stochasticwaveformestimationfundamental}. Since we require many copies of the unknown quantum state, we focus on long-lived signals. (Transient signals require a Bayesian analysis instead, cf.\ Ref.~\cite{bayesianpaper}.) Implementing WSS or DME-QSP using a spatial array of qubit memories is discussed for optical imaging in Refs.~\cite{de2025superresolving,mokeev2025enhancingopticalimagingquantum} and could be done analogously in time.

\textit{Conclusions.}---We have studied the canonical incoherent sensing problem relevant to quantum superresolution, noise spectroscopy, and Lindblad detection. If the noise foreground is not known \textit{a priori}, then existing techniques fail to detect the weak stochastic signal. We have shown that quantum computing can aid in this regime by performing WSS or DME-QSP. These algorithms beat full-state tomography, since their sample complexity is independent of the dimension. We consider the particular unitarily invariant incoherent sensing problems of rank, purity, and spectral gap testing. We have shown that the Rayleigh curse can be beaten for rank or purity testing and for spectral gap testing when the gap is much larger than the signal. We have discussed how these results may accelerate searches to detect exoplanets, stochastic waveforms such as gravitational waves, and Pauli noise. 
Open questions remain about the scaling and constants of WSS, particularly for low-rank states and limited working memory. Future work could also calculate the quantum composite Chernoff-Stein lemma~\cite{brandao2020adversarial,berta2021composite,lami2025solution,lami2025generalised} and possibly accelerate DME-QSP using virtual DME~\cite{wada2025state}.

\textit{Acknowledgements.}---We thank the following people for their advice: Yuki Koizumi, Ludovico Lami, Aleksandr Mokeev, Ryan O'Donnell, Angelos Pelecanos, Mahadevan Subramanian, Norah Tan, Kaito Wada, and Yat Wong. We acknowledge support from the ARO (W911NF-23-1-0077), ARO MURI (W911NF-21-1-0325), AFOSR MURI (FA9550-21-1-0209, FA9550-23-1-0338), DARPA (HR0011-24-9-0359, HR0011-24-9-0361), NSF (ERC-1941583, OMA-2137642, OSI-2326767, CCF-2312755, OSI-2426975), and the Packard Foundation (2020-71479). This material is based upon work supported by the U.S. Department of Energy, Office of Science, National Quantum Information Science Research Centers, and Advanced Scientific Computing Research (ASCR) program under contract number DE-AC02-06CH11357 as part of the InterQnet quantum networking project. This material is also based upon work supported by the U.S. Department of Energy, Office of Science, National Quantum Information Science Research Centers as part of the Q-NEXT center. T.G.\ acknowledges funding provided by the Quantum Science and Technology early-career fellowship of the Israel Council for Higher Education and ISF Grant No.\ 3302/25.

\bibliographystyle{bibliography/style.bst}
\bibliography{bibliography/bib.bib}




\section{Appendix}

We now prove the two propositions and Eq.~\ref{eq:QSP sample complexity} and discuss the support-preserving case, scaling in signal size, and Kraus operator detection.

\subsection{Proof of Proposition~\ref{prop:kl}}
Although we cannot directly use Theorem~9 of Ref.~\cite{grace2022perturbation} because the assumptions are different, we can closely follow the proof. The quantum Tsallis relative entropy is:
\begin{align}\label{eq:tsallis}
    D_q(\h\rho\|\h\sigma)=\frac{1}{1-q}(1-\trSmall{\h\rho^q\h\sigma^{1-q}})
\end{align}
which satisfies $\lim_{q\to1^{-}}D_q(\h\rho\|\h\sigma)=D(\h\rho\|\h\sigma)$. We use Theorem~7 of Ref.~\cite{grace2022perturbation} to expand $(\h\rho_n+\vartheta_0\h\Delta)^{1-q}$ and yield:
\begin{align}
    \label{eq:Thm.9 equiv technical}
    &\trSmall{\h\rho_n^q(\h\rho_n+\vartheta_0\h\Delta)^{1-q}}
    \\&= \tr{\h\rho_n^q\left(\h\rho_n^{1-q}+X + Y\right)}
    \nonumber\\&= \tr{\h\rho_n+\bmatrixByJames{\h\Lambda_{\vec\lambda_+}^q & 0 \\ 0 & 0 }X + \bmatrixByJames{\h\Lambda_{\vec\lambda_+}^q & 0 \\ 0 & 0 }Y}
    \nonumber\\&=1 + (1-q)\vartheta_0\trSmall{\h\Delta_\parallel} + \order{\normSmall{\vartheta_0\h\Delta}^{2}}
    \nonumber,
\end{align}
where $\h\Lambda_{\vec\lambda_+}=(1-\h\Pi_\perp)\h\rho_n(1-\h\Pi_\perp)$, $\h\Delta_\parallel=(1-\h\Pi_\perp)\h\Delta(1-\h\Pi_\perp)$, $\h\Delta_\times=(1-\h\Pi_\perp)\h\Delta\h\Pi_\perp$, and:
\begin{align}
    X &= [f_{x\mapsto x^{1-q}},\vec\lambda]^{[1,0]} \circ_H \bmatrixByJames{\vartheta_0\h\Delta_\parallel & \vartheta_0\h\Delta_\times \\ \vartheta_0\h\Delta_\times^\dagger & (\vartheta_0\h\Delta_\perp)^{1-q}}\\
    Y &= \bmatrixByJames{\order{\normSmall{\vartheta_0\h\Delta}^2} & \order{\normSmall{\vartheta_0\h\Delta}^{2-q}} \\ \order{\normSmall{\vartheta_0\h\Delta}^{2-q}} & \order{\normSmall{\vartheta_0\h\Delta}^{2-q}}}\nonumber,
\end{align}
where $\circ_H$ is the Hadamard product and $[f_{x\mapsto x^{1-q}},\vec\lambda]^{[1,0]}$ is the extended first divided difference of $x^{1-q}$ at $\vec\lambda$: 
\begin{align}
[f_{x\mapsto x^{1-q}},\vec\lambda]^{[1,0]}_{j,k} = 
     \left\{\begin{array}{ll}
    \frac{\lambda_j^{1-q}-\lambda_k^{1-q}}{\lambda_j-\lambda_k} & \text{if}\;\lambda_j\neq\lambda_k\\
    (1-q)\lambda_j^{-q} & \text{if}\;\lambda_j=\lambda_k>0\\
    1 & \text{if}\;\lambda_j=\lambda_k=0
    \end{array}\right..
\end{align}
The block matrix form for the divided difference is thus:
\begin{align}
    [f_{x\mapsto x^{1-q}},\vec\lambda]^{[1,0]} = \begin{pmatrix}
        [f_{x\mapsto x^{1-q}},\vec\lambda_+]^{[1]} & (\lambda_j^{-q})_{jk} \\
        (\lambda_k^{-q})_{jk} & J
    \end{pmatrix},
\end{align}
where $J$ is the all-ones matrix, $(\lambda_j^{-q})_{jk}$ is a matrix with constant rows, $(\lambda_k^{-q})_{jk}$ is a matrix with constant columns, and the unextended divided difference is:
\begin{align}\label{eq:unextended divided difference}
[f_{x\mapsto x^{1-q}},\vec\lambda_+]^{[1]}_{j,k} = \left\{
    \begin{array}{ll}
    \frac{\lambda_j^{1-q}-\lambda_k^{1-q}}{\lambda_j-\lambda_k} & \text{if}\;\lambda_j\neq\lambda_k\\
    (1-q)\lambda_j^{-q} & \text{if}\;\lambda_j=\lambda_k
    \end{array}\right..
\end{align}
Putting it all together, the last equality in Eq.~\ref{eq:Thm.9 equiv technical} thus holds by direct computation of the trace of the second term involving $X$, which we omit for brevity. 

The quantum Tsallis relative entropy in Eq.~\ref{eq:tsallis} is thus:
\begin{align}
    D_q(\h\rho_n\|\h\rho_n+\vartheta_0\h\Delta)
    = \vartheta_0\,\trSmall{\h\Delta_\perp} + \order{\normSmall{\vartheta_0\h\Delta}^{2}}.
\end{align}
The divergence equals the same as $q\to1^{-}$, as required. 
\qed

\subsection{Support-preserving case}
We focus on the support-extending case in the main text, since the Rayleigh curse can be avoided. We now discuss the support-preserving case where it cannot.

Let $\h\rho_n=\sum_j\lambda_j\proj{j}$ be a state on a $d$-dimensional Hilbert space $\mathcal{H}=\mathcal{H}_\parallel\oplus\mathcal{H}_\perp$ with $\text{supp}(\h\rho_n)=\mathcal{H}_\parallel$. Consider a support-preserving perturbation $\vartheta_0\h\Delta$ with $\text{supp}(\h\rho_n+\vartheta_0\h\Delta)=\mathcal{H}_\parallel$. By Theorem~3 of Ref.~\cite{grace2022perturbation}, the divergence is then given by:
\begin{align}\label{eq:Grace+23 Thm.3}
    D(\h\rho_n\|\h\rho_n+\vartheta_0\h\Delta) &=
    \frac{1}{2}\trSmall{-\vartheta_0\h\Delta L_{\log(x)}(\h\rho_n,-\vartheta_0\h\Delta)}\\&+\order{\normSmall{\vartheta_0\h\Delta}^3},
\end{align}
where the Fr\'echet derivative of the logarithm is given by 
\begin{align}
    L_{\log(x)}(\h\rho_n,-\vartheta_0\h\Delta)
    &= -\vartheta_0\sum_{j,k}[\log(x),\vec\lambda_+]^{[1]}_{j,k}\braopket{j}{\h\Delta}{k} \ket{j}\bra{k},
\end{align}
and the first divided difference of $\log(x)$ is defined as:
\begin{align}
    [\log(x),\vec\lambda_+]^{[1]}_{j,k}
    &= \left\{
    \begin{array}{ll}
    \frac{\log(\lambda_j)-\log(\lambda_k)}{\lambda_j-\lambda_k} & \text{if}\;\lambda_j\neq\lambda_k\\
    \lambda_j^{-1} & \text{if}\;\lambda_j=\lambda_k
    \end{array}\right..
\end{align}
The divergence in Eq.~\ref{eq:Grace+23 Thm.3} is thus
\begin{align}
    D&= \frac{1}{2}\vartheta_0^2\sum_{j,k}[\log(x),\vec\lambda_+]^{[1]}_{j,k}
    \absSmall{\braopket{k}{\h\Delta}{l}}^2+\order{\normSmall{\vartheta_0\h\Delta}^3},
\end{align}
which equals the usual Taylor expansion factor of $\frac{1}{2}\vartheta_0^2$ times the Kubo-Mori-Bogoliubov quantum Fisher information of $\h\rho(\vartheta)$ since $\partial_\vartheta\h\rho(\vartheta)=\h\Delta$~\cite{Hayashi_2002}. This is not the symmetric logarithmic derivative quantum Fisher information which appears in the quantum Cram\'er-Rao bound. The support-preserving case thus suffers the Rayleigh curse since the divergence scales quadratically. 

\subsection{Scaling in signal size}
We now distinguish the $M\sim1/\theta_0$ scaling in Eq.~\ref{eq:quantum Chernoff-Stein, orthogonal} from the Heisenberg limit. Consider a Bernoulli random variable with parameter $n+\theta$ with noise $n$ and signal $\theta$. The divergence in Eq.~\ref{eq:composite Chernoff-Stein} for testing between $H_0:\theta=0$ and $H_1:\theta=\theta_0$ is then:
\begin{align}\label{eq:DvE KL divergence}
    D(p\|q)=(1-n)\log\left(\frac{1-n}{1-n-\theta_0}\right)+n\log\left(\frac{n}{n+\theta_0}\right).
\end{align}
If the signal dominates the noise $n\ll\theta_0\ll1$, then the divergence scales linearly $D(p\|q)\approx\theta_0+n \log(n/\theta_0)$ and $M\sim1/\theta_0$ as in Eq.~\ref{eq:quantum Chernoff-Stein, orthogonal}. Whereas, if the noise is dominant $\theta_0\ll n$, we need to distinguish relatively small changes in the parameter and the divergence scales quadratically $D(p\|q)\approx\theta_0^2/[2n(1-n)]$ and $M\sim1/\theta_0^2$ such that the Rayleigh curse arises~\cite{oh2021quantum}. The same scaling of $M$ with $\theta_0$ in these two regimes can be shown from the Cram\'er-Rao bound on the signal-to-noise ratio:
\begin{align}
    \text{SNR} \leq \theta_0\sqrt{M\mathcal{I}_C(\theta_0)}, \quad \mathcal{I}_C(\theta_0)=\frac{1}{(n+\theta_0)(1-n-\theta_0)},
\end{align}
where we assume that $n$ is known. (The Fisher information matrix is degenerate if the noise $n$ is unknown, which is another reason that we use hypothesis testing.) The difference between the two regimes is thus classical and should not be confused with the difference between the standard quantum limit and Heisenberg limit, which is quantum~\cite{pezze2018quantum}. In particular, we do not optimise over the initial quantum state of some channel.

\subsection{Proof of Proposition~\ref{prop:wss}}
Although we cannot directly use Lemma~20 of Ref.~\cite{montanaro2013survey} because the assumptions of property testing are different, we can again closely follow the proof. Let $\h O$ be a binary projective measurement operator corresponding to the hypothesis test in the sense that its expectation value $\trSmall{\h O\h\rho^{\otimes M}}$ gives the probability to choose $\h\rho\in\mathcal{S}_0$ over $\h\rho\in\mathcal{S}_1$. We want to prove optimality of WSS in the minimax sense, i.e.\ that WSS can achieve equal or better worst-case type 1 and type 2 errors than $\h O$. Thus, let $\h\rho_0\in\mathcal{S}_0$ and $\h\rho_1\in\mathcal{S}_1$ be the states with the worst probability of acceptance/rejection in their respective hypotheses, i.e.\ $\trSmall{\h O\h\rho_0^{\otimes M}}\leq\trSmall{\h O\h\sigma^{\otimes M}}$ for all $\h\sigma\in\mathcal{S}_0$ and $\trSmall{\h O\h\rho_1^{\otimes M}}\geq\trSmall{\h O\h\varsigma^{\otimes M}}$ for all $\h\varsigma\in\mathcal{S}_1$. This worst-case analysis provides guarantees on the type 1 and type 2 errors given any states from the hypotheses. 

By the permutation invariance of $\h\rho^{\otimes M}$, we have that:
\begin{align}
    \trSmall{\h O\h\rho^{\otimes M}}
    = \frac{1}{M!}\sum_{\pi\in\text{S}(M)} \trSmall{\h O\h U_\pi\h\rho^{\otimes M}\h U_\pi^\dagger}
    = \trSmall{\hat{\overline{O}}\h\rho^{\otimes M}}
\end{align}
where $\hat{\overline{O}}:=\frac{1}{M!}\sum_{\pi\in\text{S}(M)} \h U_\pi^\dagger\h O\h U_\pi$ and $\hat{U}_{\pi}$ is a permutation. By the unitary invariance of $\mathcal{S}_0$, we then have that:
\begin{align}
    \trSmall{\hat{\overline{O}}\h\rho_0^{\otimes M}}
    \leq \inth{\h U}\trSmall{\hat{\overline{O}}(\h U\h\rho_0\h U^\dagger)^{\otimes M}}
    =\trSmall{\hat{\overline{\overline{O}}}\h\rho_0^{\otimes M}}
\end{align}
where $\text{d}\h U$ is the Haar measure on the unitary group $\text{U}(d)$ and $\hat{\overline{\overline{O}}}:=\inth{\h U}(\h U^\dagger)^{\otimes M}\hat{\overline{O}}\h U^{\otimes M}$. We therefore have that $\trSmall{\h O\h\rho_0^{\otimes M}}\leq\trSmall{\hat{\overline{\overline{O}}}\h\rho_0^{\otimes M}}$ and, similarly, that $\trSmall{\h O\h\rho_1^{\otimes M}}\geq\trSmall{\hat{\overline{\overline{O}}}\h\rho_1^{\otimes M}}$. This shows that measuring $\hat{\overline{\overline{O}}}$ achieves equal or better worst-case errors than $\h O$. By the Schur-Weyl duality, since $\hat{\overline{\overline{O}}}$ commutes with permutations and local unitaries, there must exist coefficients $0\leq c_{\vec \lambda}\leq 1$ and WSS projectors $\h \Pi_{\vec \lambda}$ such that $\hat{\overline{\overline{O}}}=\sum_{\vec \lambda}c_{\vec \lambda}\h \Pi_{\vec \lambda}$~\cite{harrow2005applications,christandl2006structure,montanaro2013survey}. Therefore, measuring $\hat{\overline{\overline{O}}}$ is equivalent to performing WSS and accepting with probability $c_{\vec \lambda}$ given outcome $\vec \lambda$. This shows that WSS is optimal with respect to the worst-case errors. \qed

\subsection{Proof of Eq.~\ref{eq:QSP sample complexity}}
We now prove Eq.~\ref{eq:QSP sample complexity}. The action of DME-QSP on an eigenvector of $\h\rho(\theta)=\sum_{j=0}^{r-1} \lambda_j\proj{j}$ is~\cite{low2017quantum,martyn2021grand,low2017hamiltonian,low2017optimal,motlagh2024generalized,wang2024resource}: 
\begin{align}\label{eq:QSP} 
    \mathcal{C}[\ket{j}\otimes\ket{0}]=\ket{j}\otimes\left[f(\lambda_jx)\ket{0}+g(\lambda_jx)\ket{1}\right]+\order{\varepsilon},
\end{align}
where the filters are trigonometric polynomials satisfying $\abs{f(y)}^2+\abs{g(y)}^2=1$. Outside the spectral gap, the filters are at least $\delta$ or $1-\delta$ up to a phase. The final state is:
\begin{align}\label{eq:QSR QSP}
    \mathcal{C}[\h\rho(\theta)\otimes\proj{0}]
    &= (1-\delta)\mathcal{E}[\h\rho(\theta)\otimes\proj{0}]+ \order{\varepsilon}
    \\&+ \delta\big[
        (1-\theta)\h\rho_n\otimes\proj{1} 
        + \theta\h\rho_s\otimes\proj{0} 
    \big] \nonumber,
\end{align}
where we drop cross terms like $\ket{0}\bra{1}$ that do not affect the measurement.
The resulting Bernoulli random variable has an unknown background $\delta+\order{\varepsilon}$ with a range of possible values $(\delta+\mu_\varepsilon-\sigma_\varepsilon,\delta+\mu_\varepsilon+\sigma_\varepsilon)$ for some $\mu_\varepsilon,\sigma_\varepsilon$. The error exponent in Eq.~\ref{eq:composite Chernoff-Stein} for $\sigma_\varepsilon\ll\delta+\mu_\varepsilon\ll\theta_0$ is:
\begin{align}\label{eq:error exponent after QSP, composite}
    \inf_{p\in \mathcal{P},q\in \mathcal{Q}}D(p\|q) 
    &\approx \theta_0+(\delta+\mu_\varepsilon +\sigma_\varepsilon ) \log \left(\frac{\delta+\mu_\varepsilon }{\theta_0}\right) -\sigma_\varepsilon,
\end{align}
which implies that $M_\text{ber}\approx\log(1/\beta)/\theta_0$, as required. 
\qed

\subsection{Kraus operator detection}
We now generalise the Lindblad detection results to any Kraus operators. A quantum channel $\mathcal{E}$ has the canonical Kraus representation $\mathcal{E}(\h\rho_i) = \sum_{j=0}^{N+S} \h K_j \h\rho_i \h K_j^\dagger$, where the Kraus operators are orthogonal $\trSmall{\h K_j^\dagger \h K_k}\propto\delta_{jk}$ and normalised $\sum_{j=0}^{N+S} \h K_j^\dagger \h K_j = \h I$. This has the form of Eq.~\ref{eq:state} if we want to sense $S$ weak Kraus operator terms in the presence of $N+1$ other terms given:
\begin{align}\label{eq:Kraus states}
    \h\rho_n = \frac{\sum_{j=0}^{N} \h K_j \h\rho_i \h K_j^\dagger}{1-\vartheta}, \quad
    \h\rho_s = \frac{\sum_{j=N+1}^{N+S} \h K_j \h\rho_i \h K_j^\dagger}{\vartheta},
\end{align}
where the parameter is $\vartheta = \sum_{j=N+1}^{N+S} \trSmall{\h K_j \h\rho_i \h K_j^\dagger}$. By Proposition~\ref{prop:kl}, only the orthogonal part of the signal from the noise contributes to the divergence, and the reduced signal is $\theta=\sum_{j=N+1}^{N+S} \trSmall{\h\Pi_\perp\h K_j \h\rho_i \h K_j^\dagger}$.

\end{document}